\documentclass[twocolumn,showpacs,preprintnumbers,amsmath,amssymb,floatfix,prd]{revtex4}
\usepackage{epsfig}
\usepackage{graphicx,pstricks}
\begin{document}
\thispagestyle{empty}
\title{$D$ mesons at finite temperature and density in the PNJL model}
\author{D. Blaschke}
\email{blaschke@ift.uni.wroc.pl}
\affiliation{Institute for Theoretical Physics, University of Wroclaw,
50-204 Wroclaw, Poland}
\affiliation{Bogoliubov  Laboratory of Theoretical Physics, JINR Dubna,
141980  Dubna, Russia}
\author{P. Costa}
\email{pcosta@teor.fis.uc.pt}
\affiliation{Departamento de F\'isica, Universidade de Coimbra, 
3004516 Coimbra, Portugal}
\affiliation{E.S.T.G., Instituto Polit\'ecnico de Leiria,
Morro do Lena-Alto do Vieiro, 2411-901 Leiria, Portugal}
\author{Yu.L. Kalinovsky}
\email{kalinov@jinr.ru}
\affiliation{Laboratory for Information Technologies, JINR Dubna,
141980  Dubna, Russia}
\begin{abstract}
We study $D$-meson resonances in hot, dense quark matter within the NJL model
and its Polyakov-loop extension.
We show that the mass splitting between $D^+$ and $D^-$ mesons is moderate,
not in excess of $100$ MeV. When the decay channel into quasifree quarks opens 
(Mott effect) at densities above twice saturation density, the decay width 
reaches rapidly the value of 200 MeV which entails a spectral broadening 
sufficient to open $J/\psi$ dissociation processes. 
Contrary to results from hadronic mean-field theories, the chiral quark 
model does not support the scenario of a dropping $D$-meson masses so that
scenarios for $J/\psi$ dissociation by quark rearrangement built on the 
lowering of the threshold for this process in a hot and dense
medium have to be reconsidered and should account for the spectral broadening.
\end{abstract}
\preprint{JINR E2-xxx 2011}
\date{\today}
\maketitle

% 11.30.Qc Spontaneous and radiative symmetry breaking
% 12.39.-x Phenomenological quark models
% 11.15.Ex Spontaneous breaking of gauge symmetries
% 12.38.Aw General properties of QCD
% 12.38.Mh Quark-gluon plasma
% 11.10.St Bound and unstable states; Bethe-Salpeter equations
% 12.38.-t Quantum chromodynamics
% 12.38.Lg Other nonperturbative calculations
% 26.60.+c Nuclear matter aspects of neutron stars
% 21.65.+f Nuclear matter

\section{Introduction}
The modification of the $D$ meson properties (masses and widths) in hot, dense
matter has consequences for scenarios of $J/\psi$ suppression, e.g., by the 
processes of the type 
\begin{equation}
\label{flip}
J/\psi + \pi,\rho \to D^* + \bar D ~~,~~ J/\psi + N \to \Lambda_c + \bar D~~,
\end{equation}
which couple hidden charm to open charm states and thus lead to a dissociation
of charm in the medium \cite{Ropke:1988bx}, see 
Refs.~\cite{Kharzeev:1994ae,Martins:1994hd,Matinyan:1998cb,Wong:1999zb} for 
early controversial estimates of the cross sections of such processes.
The reverse process \cite{Ko:1998fs,BraunMunzinger:1999gs,Thews:2000rj}
of charmonium regeneration by open charm recombination should play a dominant 
role for $J/\psi$ production at LHC \cite{Satz:2006kba,Andronic:2007bi}
where charm is abundant in the medium.

Since either dropping masses \cite{Sibirtsev:1999jr,Fuchs:2004fh}
or increasing widths \cite{Burau:2000pn,Blaschke:2003ji} of the $D$ mesons
in a hot and dense medium can lead to a lowering of the reaction threshold 
and thus to a strong increase of the rate for the processes (\ref{flip}), 
both effects may contribute to an explanation of the anomalous $J/\psi$ 
suppression found in the NA50 experiment 
\cite{Gonin:1996wn,Abreu:1997jh,Alessandro:2004ap} and 
subsequently confirmed by NA60 \cite{Arnaldi:2006ee,Arnaldi:2007zz}
and PHENIX \cite{Adare:2006nq}. For a recent review, see \cite{Rapp:2008tf}.

In contrast to results from a relativistic mean-field model of $D$ mesons
in nuclear matter which predicts a strong downwards shift of $D^+$ meson masses
due to the renormalization with a scalar mean field 
\cite{Sibirtsev:1999js}, the consideration of the quark substructure of these
mesons leads to qualitatively different behavior.
As we will show in this work on the basis of a chiral quark model of the 
Nambu-Jona-Lasinio (NJL) type and its Polyakov-loop extension (PNJL), 
the Pauli blocking effect in the Bethe-Salpeter equation for the $D$-mesons 
largely compensates the dropping masses of their quark constituents. 
As a result, $D$-meson masses do not drop but stay almost constant
or rather increase with increasing density (and temperature) of the matter.
Their decay width, however, increases rapidly and reaches values which allow
for a subthreshold quark rearrangement dissociation of $J/\psi$. 
Therefore, scenarios for $J/\psi$ suppression built on the quark rearrangement
reaction (\ref{flip}) have to be reconsidered.  
As has been demonstrated in \cite{vanHees:2004gq,vanHees:2007me}, a sufficient 
width of $D$ mesonic correlations in the quark plasma is essential for 
understanding charm thermalisation and diffusion in RHIC experiments, see 
\cite{Rapp:2009my} %\cite{Rapp:2008qc} 
for a review. 

It is interesting to note that a recent self-consistent coupled channel 
approach for $D$ mesons in hot, dense nuclear matter 
\cite{Tolos:2007vh,Tolos:2009nn}
supports the picture of a spectral broadening with a negligible mass 
shift up to temperatures $T=150$ MeV and densities $n=2~n_0$ with 
$n_0=0.16$ fm$^{-3}$ being the nucleon density of nuclear matter at saturation. 
Therefore, it seems likely that a quark hadron duality similar to that
discussed for low-mass dilepton production \cite{Rapp:1999if}
can be observed also in the $D$ meson channel at the deconfinement transition. 
The spectral broadening of $D$ mesons rather than their mass shift has been 
suggested for an explanation of anomalous $J/\psi$ suppression in 
\cite{Burau:2000pn,Blaschke:2003ji}.
This question, however, awaits a thorough investigation.

In the present note, we investigate the extension of previous exploratory
calculations of $D$ mesonic correlations in quark matter, based on the NJL 
model \cite{Gottfried:1992bz,Blaschke:2003ji} to the domain of finite baryon 
densities which will become accessible in the CBM experiment at FAIR.
Of particular interest will be the question whether the suggested mass 
splitting of $D$ meson states \cite{Weise:2001wv} will be observable or rather 
washed out by spectral broadening.
Furthermore, a strong isospin dependence of the $D$ meson broadening could 
result in observable signatures, possibly relevant for quark-gluon plasma 
diagnostics.

Chiral dynamics has been applied successfully not only in the light quark 
sector but also especially for the investigation of heavy-light pseudoscalar
meson properties.
This has been most impressively demonstrated within the 
Dyson-Schwinger equation approach in Ref.~\cite{Ivanov:1998ms} which 
reproduced a complete set of heavy meson observables.  
A particular feature of these systems is the heavy quark symmetry which 
allows to separate the physics of the heavy and the light quark components 
from each other and absorb the chiral dynamics of the latter into the 
universal Isgur-Wise function \cite{Isgur:1989vq}. 
For more details, see the reviews on heavy quark effective theory 
\cite{Neubert:1993mb,Mannel:1997ky,Grozin:2004yc}.
It is interesting to note that already the chiral dynamics encoded in the
rather schematic NJL model reproduces features like heavy-quark symmetry and 
Isgur-Wise function \cite{Ebert:1994tv} when extended to the heavy quark 
sector. 
These properties remain unaffected when confining properties are mimicked in 
the heavy-quark extended NJL model by an infrared cutoff procedure 
\cite{Ebert:1996vx} or by a confining interquark potential in a relativistic 
potential model of heavy mesons \cite{Ebert:1997nk}. 

$D$ meson properties as reported in our study come from a simultaneous
solution of two types of nonperturbative equations:
the gap equations for the quark masses and the Bethe-Salpeter equations for
the meson masses. 
The results for the dependence of pseudoscalar meson masses $M_P$ on the 
current quark masses $m_q$ of the heavier quark in the meson
nicely reproduce the transition from a Gell-Mann--Oakes--Renner like behaviour
for pion and kaon ($M_P \sim \sqrt{m_q}$) to the additive quark model like 
behaviour ($M_P \sim m_q$) for $D$ and $B$ mesons. 
These results are then readily generalized from the vacuum to finite 
temperature and chemical potential within the Matsubara formalism.

\section{Model and formalism}

We employ a four - flavor model with NJL - type interaction kernel
as a straightforward generalization of recent work on the SU$_f$(3) scalar and 
pseudoscalar meson spectrum \cite{Pedro,Costa:2003uu,Costa:2005cz} 
developed on the basis of Ref.~\cite{Rehberg:1995kh} and its 
generalization by coupling to the Polyakov loop 
\cite{Hansen:2006ee,Costa:2008dp},
\begin{eqnarray}
\label{lagr}
{\cal L} &=& \bar{q} \left( i \gamma^\mu D_\mu + \hat{m}\right) q +
G_S \sum_{a=0}^{15} \left[ \left( \bar{q} \lambda^a q\right)^2+
 \left( \bar{q} i \gamma_5 \lambda^a q\right)^2
\right]
\nonumber\\
&-& \mathcal{U}\left(\Phi[A],\bar\Phi[A];T\right). 
\end{eqnarray}
Here $q$ denotes the quark field with four flavors, $N_f=4$, 
$f=u,d,s,c$, and three colors, $N_c=3$; $\lambda^a$ are the flavor SU$_f$(4) 
Gell - Mann matrices ($a=0,1,2,\ldots,15$), $G_S$ is a coupling constant.
The global symmetry of the Lagrangian (\ref{lagr}) is explicitly broken by the 
current quark masses 
$\hat{m} = \mbox{diag}(m_u, m_d, m_s, m_c)$.
The covariant derivative is defined as
$D^{\mu}=\partial^\mu-i A^\mu$, with $A^\mu=\delta^{\mu}_{0}A^0$ 
(Polyakov gauge); in Euclidean notation $A^0 = -iA_4$.  
The strong coupling constant $g_s$ is absorbed in the definition of 
$A^\mu(x) = g_s {\cal A}^\mu_a(x)\frac{\lambda_a}{2}$, where 
${\cal A}^\mu_a$ is the (SU$_c$(3)) gauge field and $\lambda_a$ are the (color)
Gell-Mann matrices.

The Polyakov loop field  $\Phi$ appearing in the potential term of
(\ref{lagr}) is related to the gauge field through the gauge covariant
average of the Polyakov line~\cite{Ratti:2005jh}
\begin{equation}
\Phi(\vec x)=\left\langle \left\langle l(\vec x)\right\rangle\right\rangle
=\frac{1}{N_c}{\rm Tr}_c\left\langle \left\langle L(\vec x)
\right\rangle\right\rangle,
\label{eq:phi}
\end{equation}
where
\begin{equation}
L(\vec x) ={\cal P}\exp\left[i\int_0^\beta d\tau A_4(\vec x, \tau)\right]\,.
\label{eq:loop}
\end{equation}
Concerning the effective potential for the (complex) $\Phi$ field, we adopt 
the form and parametrization proposed in Ref.~\cite{Roessner:2006xn}.

This effective chiral field theory has the same chiral symmetry of QCD,
which is also shared by the  quark interaction terms. 
The (P)NJL model is a primer for describing the dynamical breakdown of this 
symmetry in the vacuum and its partial restoration at high
temperatures and chemical potentials. 
At the same time it provides a field-theoretic description of 
pseudoscalar meson properties which is in accordance with the low energy 
theorems (such as the Goldstone theorem) of QCD.  

In the vacuum the PNJL model with the Lagrangian (\ref{lagr}) goes over to the 
NJL one and the pseudoscalar meson properties are described in the standard way 
by analyzing the polarization operators 
\begin{eqnarray}
\Pi_{ij} (P) = i N_c \int \frac{d^4p}{(2\pi)^4}\mbox{tr}_{D}\left[ 
%{\cal P}_{ij}(p,P) \right] \\
%{\cal P}_{ij} = 
S_i (p) (i \gamma_5 )S_j (p+P)(i \gamma_5 ) \right]~,
\end{eqnarray} 
where $\mbox{tr}_{D}$ is the trace over Dirac matrices, $S_i(p)$ is the quark 
Green function with the dynamical quark mass $M_i$. 

The polarization operators can be presented in terms of two integrals
which for mesons at rest in the medium are given by 
\begin{eqnarray}
\Pi^{ij} (P_0) &=& 4 \bigl\{ \left( I_1^i+I_1^j \right) \nonumber \\
&&-\left[  P_0^2 -(M_i-M_j)^2 \right] I_2^{ij}(P_0) \bigr\}, 
\end{eqnarray} 
where 
\begin{eqnarray}\label{i1}
I_1^i &=& i N_c \int \frac{d^4p}{(2\pi)^4} \, \frac{1}{p_0^2-E_i^2}
       = \frac{N_c}{4 \pi^2} \int^{\Lambda}_0 \frac{{\tt p}^2 d {\tt p}}{E_i} ,
\end{eqnarray}
\begin{eqnarray}\label{i2}
I_2^{ij}(P_0) &=& i N_c \int \frac{d^4p}{(2\pi)^4} \, 
\frac{1}{(p_0^2-E_i^2)((p_0+P_0)^2-E_j^2)}
\nonumber \\
       &=& \frac{N_c}{4 \pi^2} \int^{\Lambda}_0 \frac{{\tt p}^2 d {\tt p}}{E_i E_j}
       \,\,\, \frac{E_i+E_j}{P_0^2-(E_i+ E_j)^2} \, ,
\end{eqnarray}
where $E_{i,j}=\sqrt{{\tt p}^2+M_{i,j}^2}$ is the quark energy.

As the Lagrangian (\ref{lagr}) defines a nonrenormalizable 
field theory, we introduce the 3 - momentum cutoff with the
parameter $\Lambda$ to regularize the integrals. 
When $P_0 > M_i+M_j$ it is necessary  to take into account
the imaginary part of the second integral.
It may be found, with help of the $i \epsilon$ --prescription
$P_0^2 \rightarrow P_0^2 - i \epsilon$, that
\begin{eqnarray}\label{ima}
I_2^{ij}(P_0)
     & =& \frac{N_c}{4 \pi^2}
       {\mathcal{P}}\int^{\Lambda}_0 \frac{{\tt p}^2 d {\tt p}}{E_i E_j}
       \,\,\frac{E_i+E_j}{P_0^2-(E_i+ E_j)^2} \nonumber \\
&&+       i \frac{N_c}{16\pi} \, \frac{{\tt p}^*}{(E_i^*+E_j^*)},
\end{eqnarray}
where ${\tt p}^*=\sqrt{(P_0^2-(M_i-M_j)^2)(P_0^2-(M_i+M_j)^2)}/2P_0$  is  
the momentum and
 $E^*_{i,j}=\sqrt{({\tt p}^*)^2+M_{i,j}^2}$  
the corresponding energy. 

The quark mass $M_i$ we find from the gap equation
\begin{eqnarray}
M_i = m_i + 16 M_i G_S I_1^i. 
\end{eqnarray}
The meson mass spectrum we obtain from the condition
\begin{eqnarray}\label{mass}
1- 2G_S  \Pi^{ij}(P_0=M_P, {\bf P}=0) =0~.
\end{eqnarray}
The pseudoscalar meson-quark-antiquark coupling constants are defined as
\begin{eqnarray}
g^{-2}_{P\bar{q}_iq_j} = - \frac{1}{2M_P} \frac{\partial}{\partial P_0} 
\left[ \Pi^{ij}(P_0) \right]{ \big|_{ P_0=M_P}}. 
\end{eqnarray}

Note that when $P_0>M_i+M_j$, then Eq.~(\ref{mass}) has to be solved in their 
complex form in order to determine the mass of the resonance $M_P$ and the 
respective decay width $\Gamma_P$. 
Thus, we assume that this equation can be written as a system of two coupled 
equations
\begin{eqnarray}
\label{BSE1}
&&-\left[ M_P^2-\frac{1}{4}\Gamma_P^2-(M_i-M_j)^2 \right] = \nonumber \\ 
&& \frac{(8G_S)^{-1}-(I_1^i+I_1^j)}{|I_2(P_0=M_P+i\epsilon)|^2}
\mbox{Re} I_2(P_0=M_P+i\epsilon),\\ 
&&-\left[ M_P \Gamma_P \right] = \nonumber \\ 
 && \frac{(8G_S)^{-1}-(I_1^i+I_1^j)}{|I_2(P_0=M_P+i\epsilon)|^2}
\mbox{Im} I_2(P_0=M_P+i\epsilon),  
\label{BSE2}
\end{eqnarray}
which have solutions of the form 
\begin{eqnarray}
P_0=M_P-i\frac{1}{2}\Gamma_P~. 
\end{eqnarray}

\begin{figure}
\includegraphics[height=.45\textwidth,width=0.45\textwidth,angle=0]{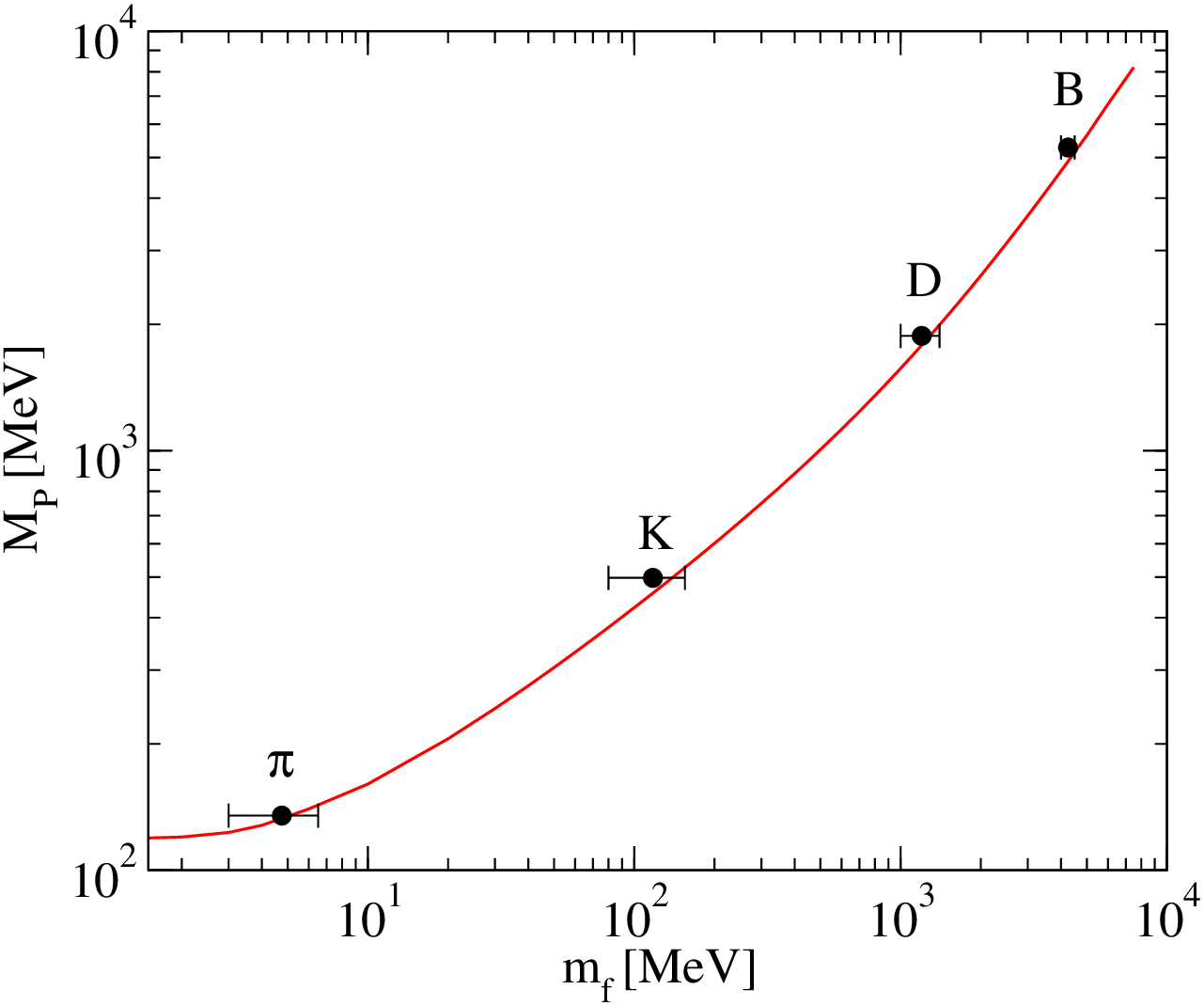}
\caption{(Color online)
The systematics of heavy-light pseudoscalar meson masses $M_P$ in dependence on
the mass $m_f$  of the heavy flavor in the bound state with a light antiquark
within the present model (red solid line) is compared to the masses of 
heavy-light pseudoscalar mesons ($\pi$, $K$, $D$, $B$) and the limits 
for the corresponding current quark masses (symbols with error bars) according 
to the particle data group \cite{PDG}. 
\label{fig:mass}
}
\end{figure}

As shown in \cite{Pedro}, the model (\ref{lagr}) succesfully describes    
meson properties in the vacuum at $T=\mu=0$. 
We use here the parametrization given in table III of 
Ref.~\cite{Grigorian:2006qe} for the case of the NJL model.
The value of the coupling constant is $G_S\Lambda^2=2.32$, the three-momentum
cutoff is at $\Lambda=602.3$ MeV, see also \cite{Sandin} for an online
tool fo the three-flavor NJL model.
The parametrization of the current-quark masses for the heavy flavors $(c, b)$
is performed with the above formulae for the masses of the corresponding 
heavy-light pseudoscalar mesons $(D,B)$.
The results are summarized in table~\ref{tab:par}.
The dependence of the heavy-light meson mass on the current mass of the 
heavier quark is shown by the solid line in Fig.~\ref{fig:mass} and provides
satisfactory agreement with the particle data group listings~\cite{PDG}.

\vspace*{0.2cm}

\begin{table}[htb]
\begin{tabular}{l|c|cc|cc}
\hline
\hline
flavor & $P$ & $M_P$ & $f_P$ & $m_f$ & $M_f$ \\
       &     & [MeV] & [MeV] & [MeV] & [MeV] \\
\hline
u,d    & $\pi$& 135.0 & 92.4 & 5.5 & 368.0 \\
s      & $K$ &  497.7 & 95.4 & 140.7 & 587.4 \\
c      & $D$ & 1869.3 & 79.6 & 1279.9 & 1828.8 \\
b      & $B$ & 5279.4 &   & 4634.8 &  \\
\hline
\hline
\end{tabular}
\caption{Results for pseudoscalar meson properties in the light, strange,  
charm and bottom sectors for the corresponding current quark masses and 
dynamically generated quark masses of the model in the vacuum at $T=\mu=0$.
\label{tab:par}}
\end{table}
\vspace*{0.2cm}

The generalization of the model (\ref{lagr}) to the case of nonzero 
temperature and density (details in \cite{Pedro}) is done within the 
imaginary time formalism by introducing the Matsubara frequencies 
$\omega_n=(2n+1)\pi T$, $n=0,\pm 1, \pm 2, \ldots$, so that 
$p_0 \longrightarrow i\omega_n + \mu$ with the chemical potential $\mu$ and the 
temperature $T$.
Instead of the integration over $p_0$, we have now to perform a sum over 
Matsubara frequencies. In the result we obtain the integrals 
\begin{eqnarray}\label{firstt}
I_1^i = - \frac{N_c}{4\pi^2} \int \frac{{\tt p}^2 d{\tt p}}{E_i} \left(
n^+_i - n^-_i
\right),
\end{eqnarray}

\begin{eqnarray}
&&I_2^{ij} (P_0,T,\mu) = \nonumber \\
&& - N_c \int \frac{d^3{\bf p}}{(2\pi)^3}
\Biggl[
\frac{1}{2E_i}
\frac{1}{(E_i+P_0)^2-E_j^2} \,\, n^+_i \nonumber \\
&& \hspace*{0cm}
 - \frac{1}{2E_i}  \frac{1}{(E_i-P_0)^2-E_j^2} \,\, n^-_i
\nonumber \\ &&
\hspace*{0cm}+ \frac{1}{2E_j}\frac{1}{(E_j-P_0)^2-E_i^2}
\,\,  n^+_j  %      \frac{1}{(p_0+P_0-E_j)}
\nonumber \\
  && \hspace*{0cm}- \frac{1}{2E_j}\frac{1}{(E_j+P_0)^2-E_i^2}
\,\,  n^-_j %  \frac{1}{(p_0+P_0+E_j)}
\Biggr], 
\end{eqnarray}
where  $n_i^{\pm}=f_\Phi(\pm E_i)$ are the generalized fermion distribution 
functions \cite{Costa:2008dp} for quarks of flavor $i$ with positive (negative) 
energies in the presence of the Polyakov loop values $\Phi$ and $\bar{\Phi}$ 
\begin{eqnarray}
\label{fermi-Pol}
f_\Phi(E)=\frac{\bar{\Phi}{\rm e}^{-\beta(E-\mu)}+2\Phi {\rm e}^{-2\beta(E-\mu)}
+{\rm e}^{-3\beta(E-\mu)}}{1+3(\bar{\Phi}+\Phi{\rm e}^{-\beta(E-\mu)})
{\rm e}^{-\beta(E-\mu)}+{\rm e}^{-3\beta(E-\mu)}}~,\nonumber\\
\end{eqnarray} 
which go over to the ordinary Fermi functions in the case of the NJL model,
where $\Phi=\bar \Phi=1$
\begin{eqnarray}\label{fermi}
         f_1(\pm E) = \frac{1}{1+ {\rm e}^{\beta (\pm E - \mu)}}~.
\end{eqnarray} 
Note that we put the chemical potential for charm quarks to zero in the 
calculations discussed below.
A small finite value might be necessary to ensure exactly vanishing net charm 
at high baryon density when, as we demonstrate below, the symmetry between 
masses of $D$ mesons with charm and those with anticharm is broken by medium 
effects.

\section{Results} 

\begin{figure}[!t]
\includegraphics[width=0.45\textwidth,angle=0]{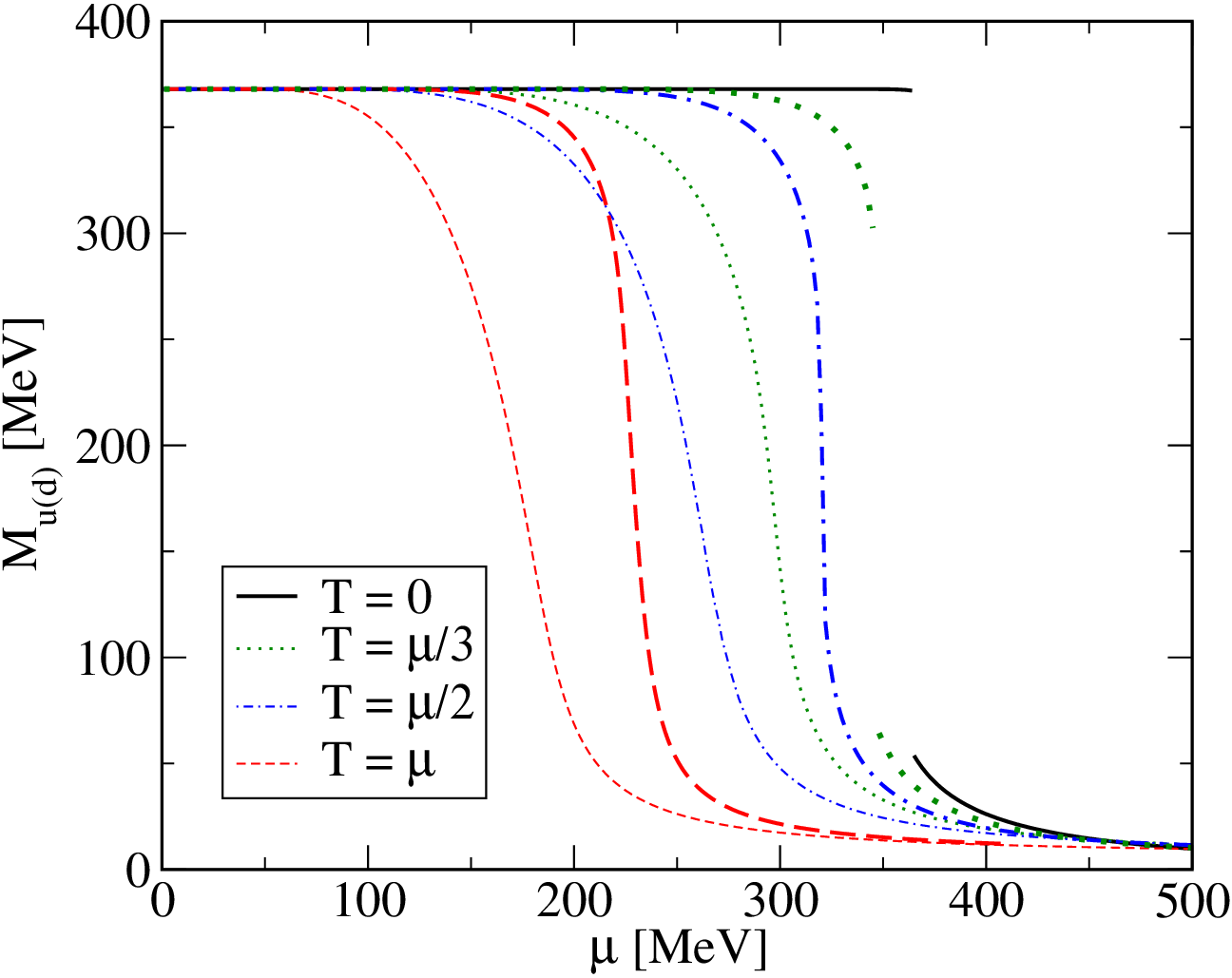}
\caption{(Color online)
Dynamically generated light quark mass $M_{u(d)}$ as a function of the quark 
chemical potential along trajectories in the quark matter phase diagram for 
$T=0$ (black solid line), $T=\mu/3$ (green dotted line), $T=\mu/2$ 
(blue dash-dotted line) and $T=\mu$ (red dashed line). 
Results of the NJL model (thin lines) are compared to those of the PNJL model. 
They coincide for $T=0$.
\label{quarks}
}
\end{figure}

\begin{figure}[!t]
\includegraphics[width=.45\textwidth]{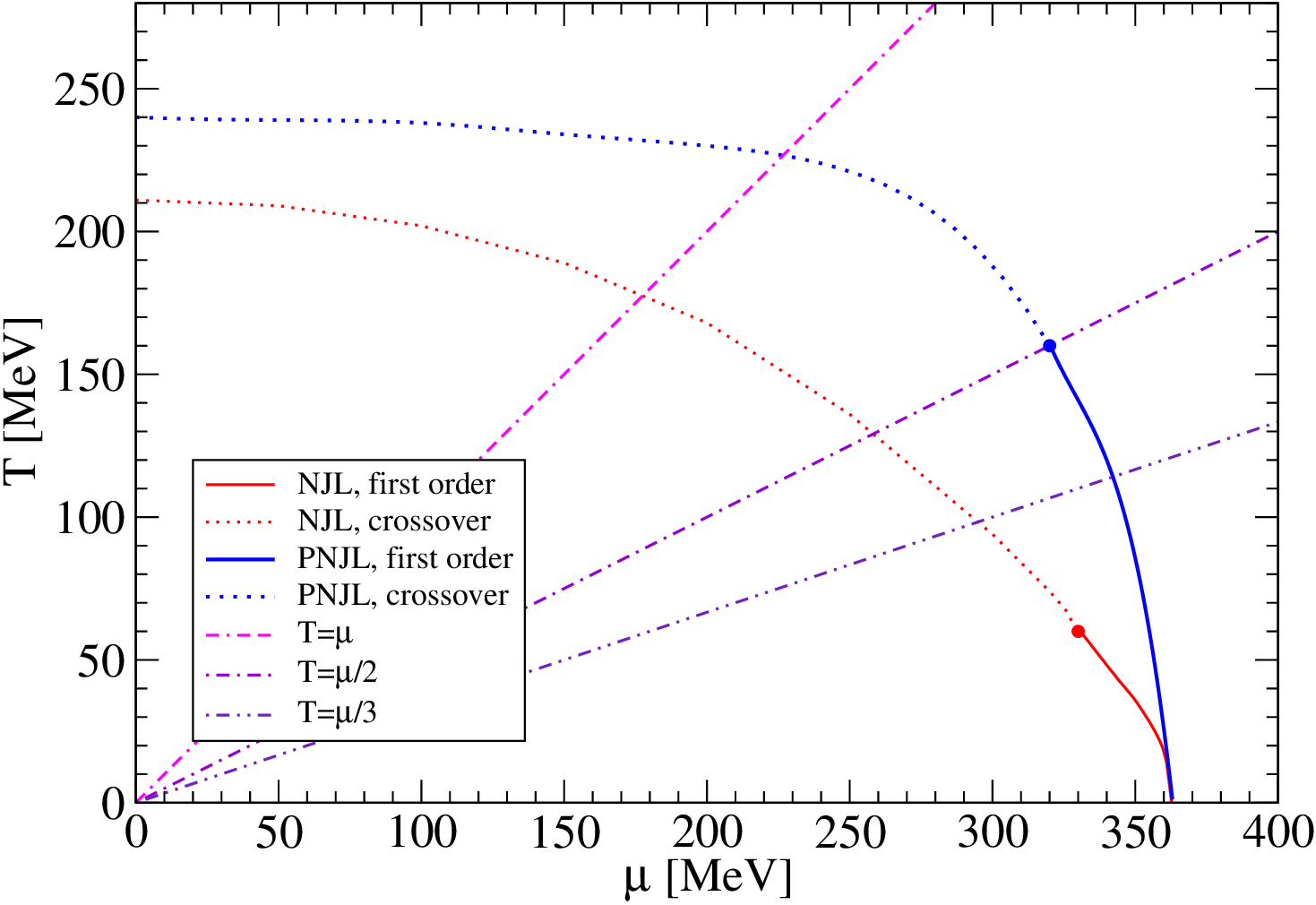}
\caption{(Color online)
(Pseudo)critical temperatures versus quark chemical potential for the 
NJL model and for the PNJL model. The first order transition lines are 
shown as solid lines, the crossover is indicated by dashed lines. 
Critical end points are denoted as full dots.
The straight lines stand for the trajectories $T=r \mu$ along which we 
investigate the $D$ meson properties: $r=1, 1/2, 1/3$.
\label{PhD}
}
\end{figure}

We have performed selfconsistent solutions of the gap equations for the 
dynamically generated light quark masses at finite temperatures and chemical 
potentials for the PNJL model and its NJL model limit within the standard
setting as summarized above.
In Fig.~\ref{quarks} we display the results for the restoration of the 
approximate chiral symmetry in the $u, d$ quark sector along trajectories
in the QCD phase diagram with a constant ratio $r=T/\mu$, where 
$r=0,~1/3,~1/2,~1$. 
The (pseudo)critical temperatures for the 
NJL model and for the PNJL model are shown in the phase diagram 
in Fig.~\ref{PhD} together with the trajectories along which we 
investigate the $D$ meson properties.

When lowering the ratio $r\to 0$, the phase transition turns from crossover 
to first order. 
The chiral restoration is a result of the phase space occupation 
(Pauli blocking) which effectively reduces the interaction strength in the gap 
equation. 

\begin{figure}[!h]
\includegraphics[height=.45\textwidth,width=0.45\textwidth,angle=0]{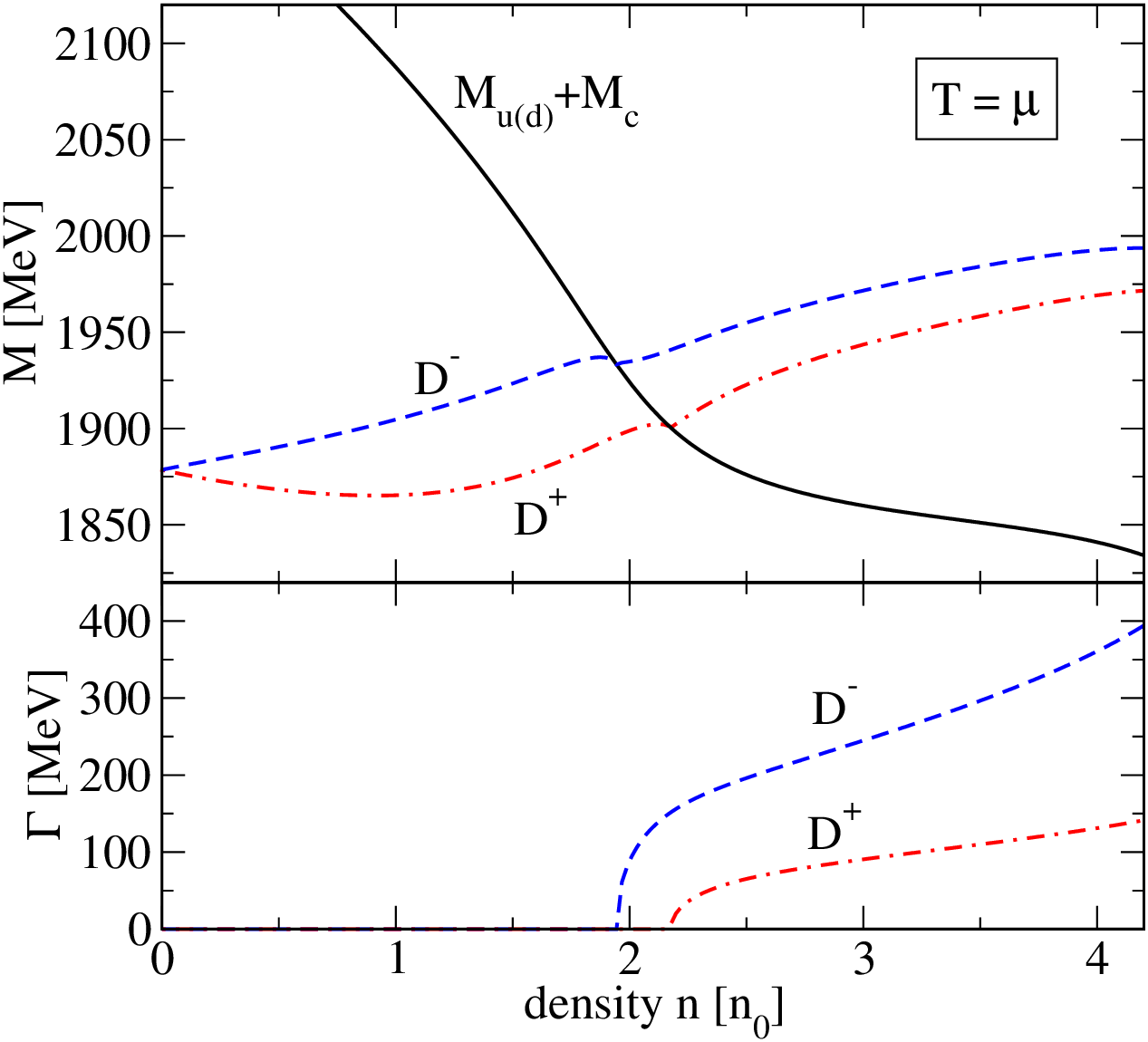}
\caption{(Color online)
Density dependence of $D$ meson masses (upper panel) and widths (lower panel)
along the trajectory  $T=\mu$ in the QCD phase diagram for the NJL model case. 
While the $D$ meson bound state masses are almost constant or moderately rise 
with increasing density the continuum threshold ($M_{u(d)}+M_c$) gets lowered
dramatically due to the chiral symmetry restoration transition.
At about twice nuclear density, the decay channel into their quark-antiquark 
constituents opens and the bound states become resonant scattering states in
the quark-antiquark continuum (Mott effect).
\label{Dmesons-Tmu}
}
\end{figure}

\begin{figure}[!h]
\includegraphics[height=.45\textwidth,width=0.45\textwidth,angle=0]{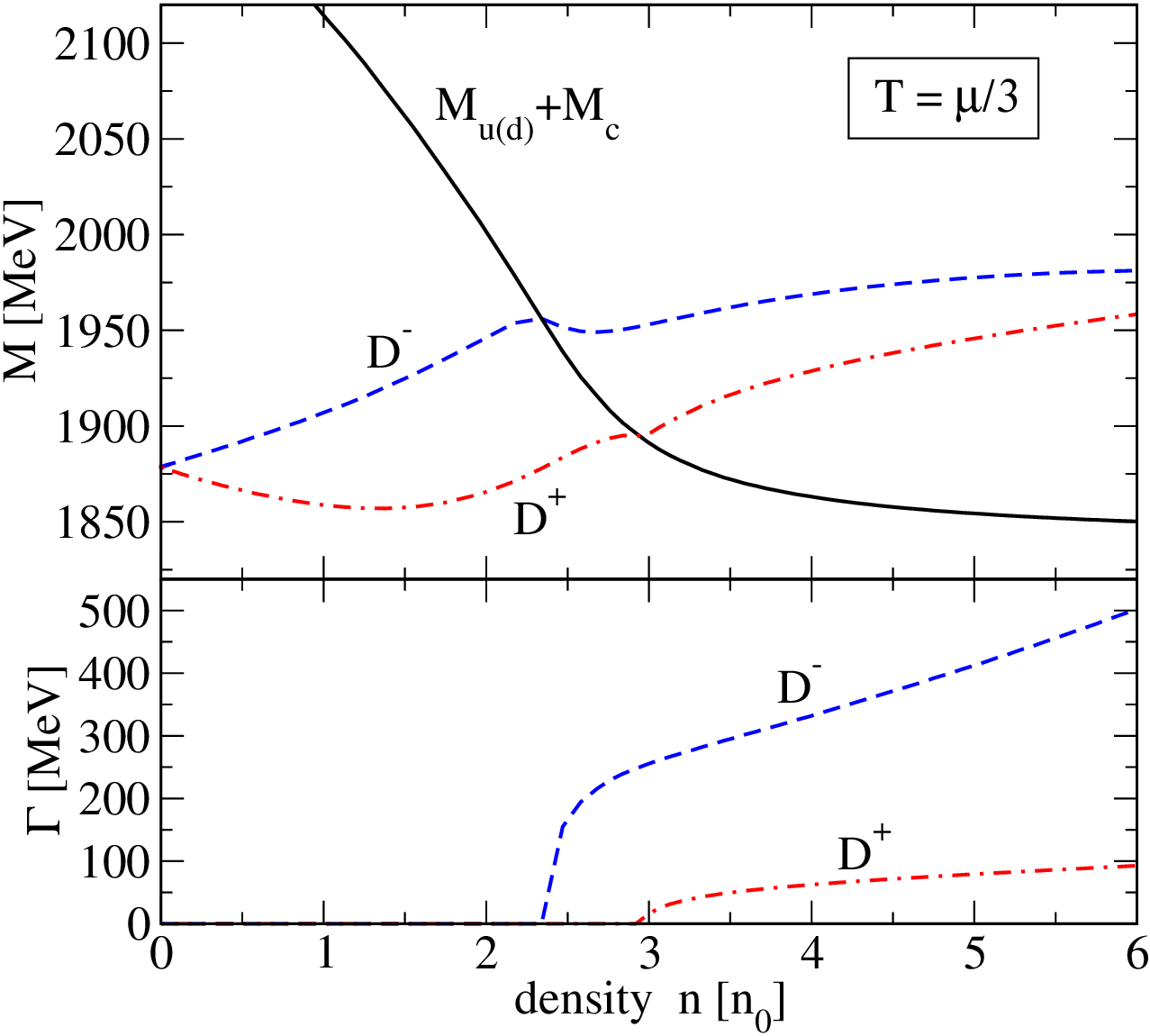}
\caption{(Color online)
Same as Fig.~\ref{Dmesons-Tmu} along the trajectory  $T=\mu/3$.
\label{Dmesons-T3mu}
}
\end{figure}

Na\"ively, one would expect that heavy-light mesons such as $D $ mesons, 
should suffer a mass reduction when embedded in a hot and dense medium, as a 
result of the strong reduction of the light quark constituent mass (the 
charm quark mass is approximately unaffected). 
The solution of the in-medium  $D^-$ meson Bethe-Salpeter equation (BSE), 
however, shows a different result, displayed in 
Figs.~\ref{Dmesons-Tmu},\ref{Dmesons-T3mu} for the NJL case.
This can be understood from inspecting the kernel of the BSE: 
The mass shift of the light quarks, which affects the energy denominators
and would lead to a lowering of the meson masses, when evaluated in free space,
gets compensated or even overcompensated by the Pauli blocking factor in the 
numerator. It is clear that $D$ mesons containing a light quark, as the
dominant quark species in a dense medium, feel a stronger Pauli blockig than 
those containing light antiquarks. Therefore, the $D^-$ mass is shifted 
upwards while the $D^+$ mass stays approximately constant.

There is an important consequence of this fact that the heavy-light continuum 
threshold is lowered while the bound state masses are not: at a critical 
density the bound states merge the continuum (Mott effect) and become unstable 
against decay into their quark constituents, as signalled by the nonvanishing 
decay widths shown in the lower panels of Figs.~\ref{Dmesons-Tmu},
\ref{Dmesons-T3mu}.
The stronger Pauli blocking for the $D^-$ mesons leads not only to a lower
Mott density, but also to a larger decay width as compared to the  $D^+$. 
From these Figures one can also read off the critical densities where the 
phenomenon of $D$ meson Mott effect could be expected for the NJL model.

\begin{figure}[htb]
\includegraphics[width=.47\textwidth,angle=0]{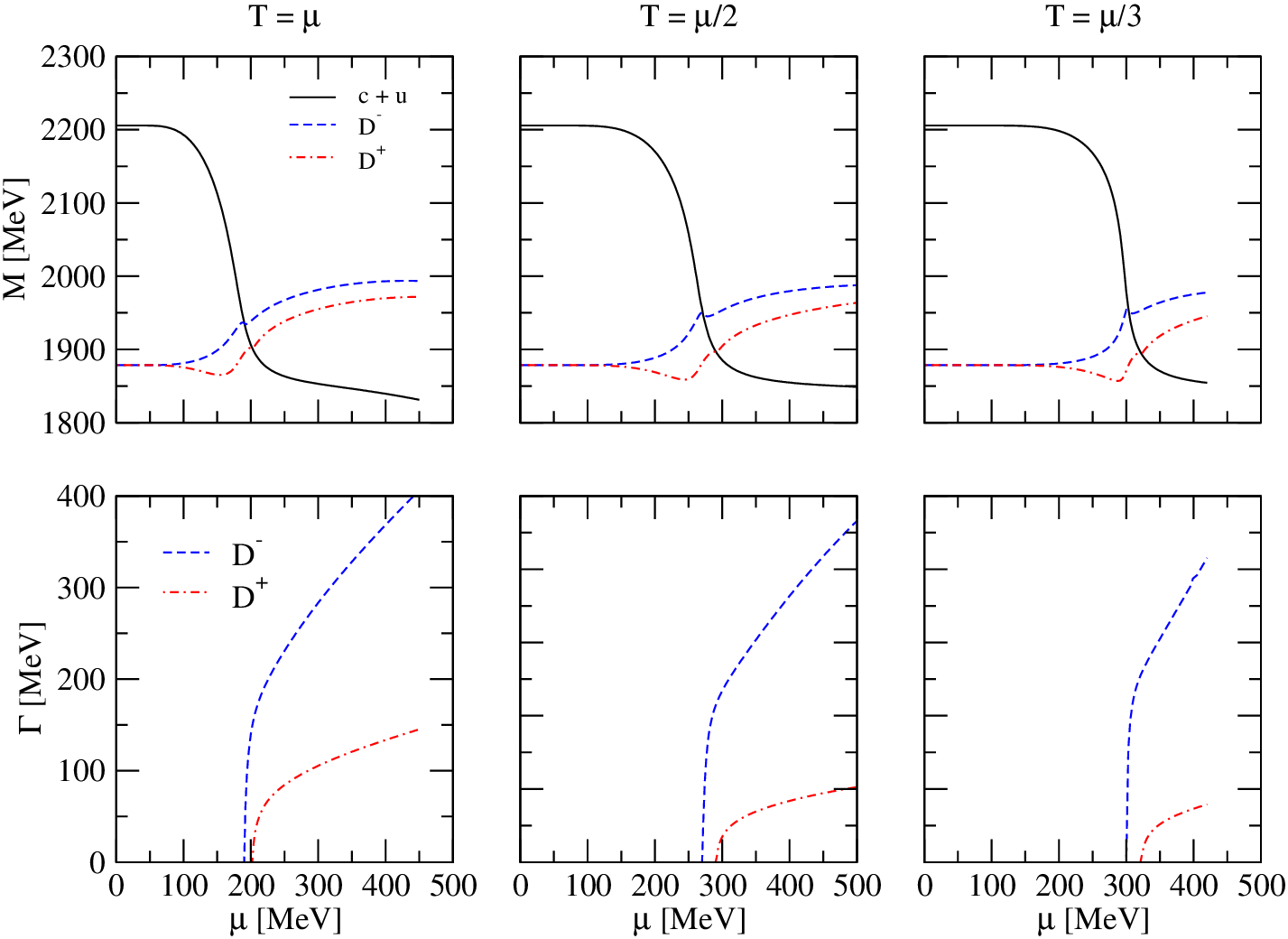}
\caption{(Color online) $D$ meson masses and widths along the trajectories
$T=\mu$, $T=\mu/2$ and $T=\mu/3$ in the QCD phase diagram for the NJL model. 
\label{Dmu_NJL}
}
\end{figure}

\begin{figure}[bht]
\includegraphics[width=.47\textwidth,angle=0]{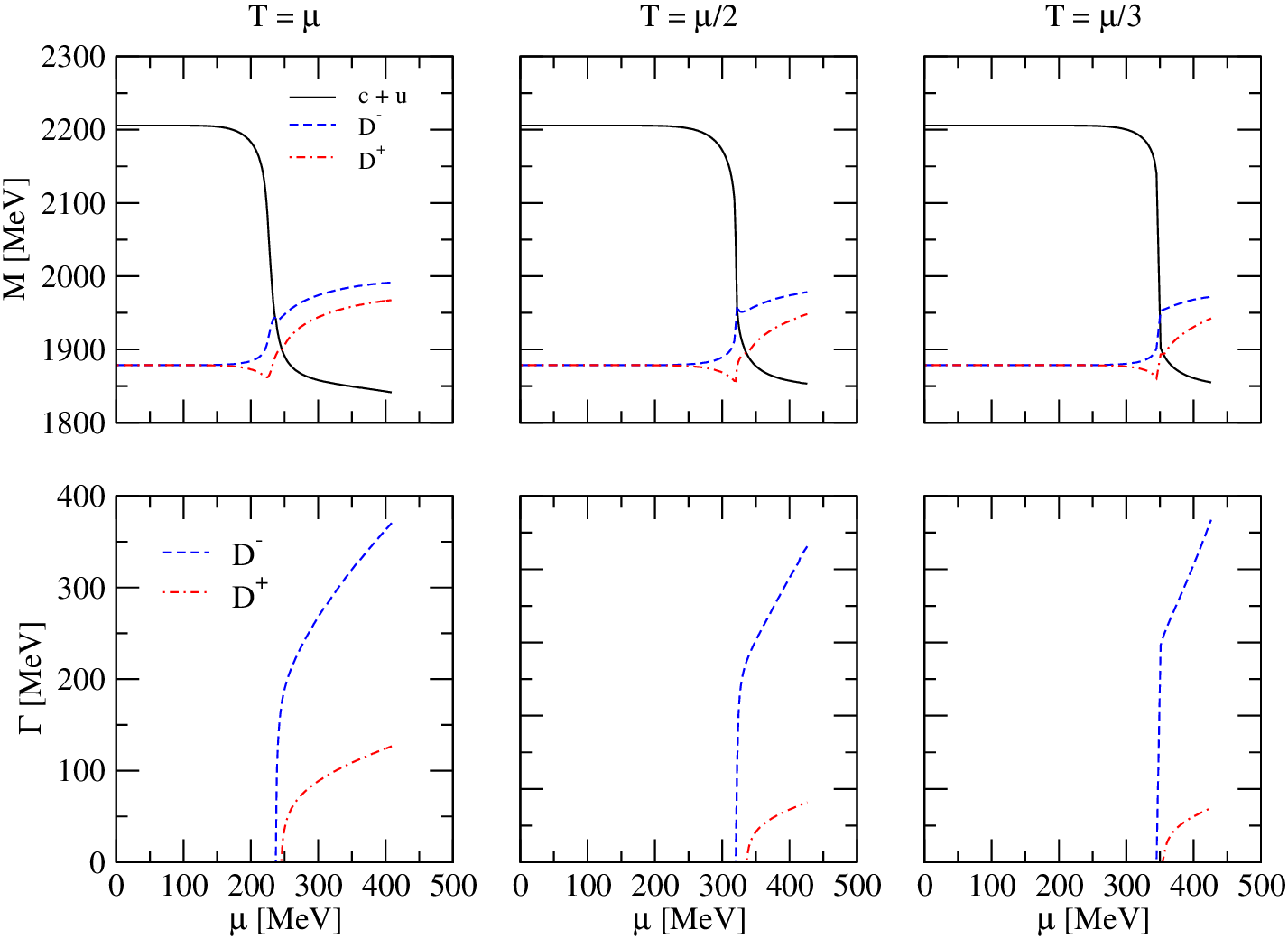}
\caption{(Color online) Same as Fig.~\ref{Dmu_NJL} for the PNJL model. 
\label{Dmu_PNJL}
}
\end{figure}

In Figs.~\ref{Dmu_NJL},\ref{Dmu_PNJL} we present results as a function of the
quark chemical potential in isospin symmetric matter for the NJL model and its
extension with coupling to the Polyakov loop.
The results are rather similar, except for the smaller width of the transition
region in the PNJL case and the different position of the critical point in 
the phase diagram. 
The higher temperature of the critical endpoint entails that along the 
trajectory for $r=1/3$ the medium undergoes a first order transition and the 
$D$ meson properties change discontinuously.
 
%\begin{widetext}
%\end{widetext}

%\begin{widetext}
%\end{widetext}

The modification of $D$ meson properties in hot and dense nuclear matter 
as reported in the present work is essentially different from the one suggested 
in Ref.~\cite{Sibirtsev:1999jr},
where a lowering of the $D$ meson mass had been conjectured with consequences 
for charmonium dissociation in heavy-ion collision experiments.
In this reference the Pauli blocking effect was neglected.
It is interesting to note that the Pauli blocking effect occurs not only on the
quark level but also on the hadronic level of description, when coupled 
channel equations for $D$ mesons in nuclear matter are solved selfconsistently
as, e.g., in Ref.~\cite{Tolos:2007vh}. 
Also in this approach the $D$ meson mass remains almost unshifted while a 
considerable spectral broadening is obtained under similar conditions of
density and temperature as considered in the present work. 

\section{Conclusions}
In the isospin-symmetric quark matter case holds
$M_{D^+}=M_{D^0}$ and $M_{D^-}=M_{\bar D^0}$.

The $D$ mesons containing light quarks ($D^-=\bar cd$, $\bar D^0=\bar cu$) 
suffer a positive mass shift due to the effective reduction of the coupling 
by Pauli blocking, since the phase space is occupied by light quarks abundant 
in the medium. 
The $D$ mesons composed of light antiquarks ($D^+=\bar dc$, $D^0=\bar uc$)
suffer no Pauli shift since there are no antiquarks in the medium.
Their mass at the Mott transition density is approximately the same as in 
vacuum.

It is interesting to compare the decay widths of the $D$ mesons into their
quark constituents. 
Due to the repulsive Pauli shift, the $D^-$ and $\bar D^0$
are clearly above the threshold already at the first order chiral phase
transition and have a non-negligible decay width, whereas the  $D^+$ and $D^0$ 
mesons are still very good resonances with negligible decay width at the 
transition.

The effect of coupling the chiral quark model to the Polyakov loop in the PNJL 
model is an effective suppression of the quark distribution functions in the 
BSE (\ref{BSE1}),(\ref{BSE2}) as long as $\Phi\ll 1$ which leads to a narrowing 
\cite{Horvatic:2010md} of the chiral transition region where medium effects on 
the D meson properties occur in the present model. 
In this region the dissociation of D mesons (Mott effect) occurs
and is signalled by an increase in the spectral width of these states. 

Summarizing the results of this model, we find no support for dropping $D$ 
meson masses in the vicinity of the chiral/deconfinement phase transition in
hot and dense matter but rather a strong spectral broadening. 
Therefore, scenarios of $J/\psi$ suppression in dense matter via dissociation 
processes like (\ref{flip}),
which are built on increasing widths \cite{Burau:2000pn,Blaschke:2003ji} for 
the $D$ mesons should be conceptually preferable over those built on dropping 
masses.

\section*{Acknowledgement}
The work of Yu.L.K was supported by DFG under grant No. BL 324/3-1 and by the 
Russian Fund for Basic Research (RFBR) under grant No. 09-01-00770a.
D.B. acknowledges partial support from the Polish Ministry of Science and 
Higher Education (MNiSW) under grant No. NN 202 2318 37 and by RFBR under
grant No. 11-02-01538-a.
The work of P.C. was supported by FCT under Project No. CERN/FP/116356/2010.

\end{document}